\documentclass[11pt,reprint,showkeys, amsmath,amssymb, aps, nofootinbib, floatfix]{revtex4-2}
\DeclareUnicodeCharacter{2212}{\textendash}
\usepackage{graphicx,subfigure}
\graphicspath{{figures/}}
\usepackage[export]{adjustbox}
\usepackage{dcolumn}
\usepackage[outdir=./epsTopdf/]{epstopdf}
\usepackage{bm}
\usepackage{physics}
\usepackage{xcolor}
\usepackage{slashed,cancel}
\usepackage{multirow}
\usepackage{float}
\usepackage{capt-of}
\usepackage{comment}
\usepackage{mathrsfs}
\usepackage{enumitem}
\usepackage[colorlinks=true,linkcolor=blue,urlcolor=blue,citecolor=blue]{hyperref}% add hypertext capabilities

\allowdisplaybreaks
\begin{document}
%\preprint{APS/123-QED}

\title{Probing anomalous quartic gauge couplings in same-sign $W$ boson scattering with polarization and spin correlation}

\author{Oscar J. P. Éboli}
\email{eboli@fma.if.usp.br}
\author{Rafiqul Rahaman}
\email{rafiqul@if.usp.br} 

\affiliation{Instituto de Física, Universidade de São Paulo, São Paulo, SP 05508-090,  Brasil}
\author{Amir Subba}
\email{amirsubba@ustc.edu.cn}
\affiliation{Wilczek Quantum Center, Shanghai Institute for Advanced Studies, Shanghai 201315, China}
\affiliation{University of Science and Technology of China, Hefei 230026, China }
%\date{\today}

\begin{abstract}
The study of quartic couplings of electroweak gauge bosons not only
provides a test of the Standard Model (SM) predictions, but also can look for signals of new physics  beyond the SM. We present a comprehensive study of anomalous quartic gauge couplings in same-sign $W^\pm W^\pm$ production via vector boson scattering at the LHC. The analysis is carried out within the framework of the SM Effective Field Theory, exploiting polarization and spin-correlation effects encoded in angular asymmetries in addition to conventional kinematic observables. We demonstrate that spin-correlation asymmetries provide sensitivity to anomalous $WWWW$ interactions that is comparable to that obtained from the transverse mass distribution of the $WW$ system. By identifying a minimal set of the most sensitive asymmetries, we show that the dominant constraints on the Wilson coefficients can be captured with a reduced number of observables. A combined analysis of angular asymmetries and kinematic information leads to  improved limits compared to either approach alone. The impact of unitarity considerations is also examined by imposing invariant-mass cut-offs on the $WW$ system, allowing us to determine unitarity-safe regions for the anomalous couplings.
\end{abstract}

\maketitle

%%%%%%%%%%%%%%%%%%%%%%%%%%%%%%%%%%%%%%%%%%%%%%%%%%%%%%%%%%%%%%%%%%%%%%
\section{Introduction}
\label{sec:intro}

The CERN Large Hadron Collider (LHC) has already accumulated
a large statistics, and it is planned to collect much more in the high
luminosity (HL-LHC) run. LHC’s large dataset allows not only precision
tests of the Standard Model (SM) but also detailed searches for new
physics beyond the SM (BSM). The SM predicts the quartic gauge-boson couplings (QGC), ergo
their study is a further test of the SM as well as a sensitive test of
new physics in case departures from the SM predictions are observed.
In order to avoid strong bounds originating from the collider studies
of triple gauge-boson couplings (TGC), here we focus on the so-called
genuine QGC operators, that is, effective operators generating QGC, but
that do not generate any TGC; for models leading to such operators
see Ref.~\cite{Godfrey:1995cd}.\smallskip

In collider experiments, QGCs contribute to the production of three
electroweak vector bosons~\cite{Belanger:1992qh, Dervan:1999as,
Eboli:2000ad,Chatrchyan:2014bza,Aad:2015uqa,Aaboud:2017tcq,
Sirunyan:2017lvq}, the exclusive production of gauge-boson
pairs~\cite{ Belanger:1992qi, Chatrchyan:2013akv,
Khachatryan:2016mud}, and the vector-boson-scattering (VBS)
production of electroweak vector boson pairs~\cite{ Belyaev:1998ih,Eboli:2000ad,Eboli:2003nq, Khachatryan:2016vif, Aaboud:2016ffv,Green:2016trm,ATLAS:2016snd,Khachatryan:2017jub,CMS:2017fhs,Sirunyan:2017fvv, Sirunyan:2019der,Sirunyan:2020tlu,CMS:2020fqz,CMS:2020gfh, Hwang:2023wad}. In these
processes, anomalous QGC (aQGC) leads to a rapid growth of the cross sections,
therefore, requiring a unitarization procedure to avoid unphysical
theoretical predictions of  cross
sections~\cite{Perez:2018kav,Almeida:2020ylr}. \smallskip

In this work, we probe aQGC in the VBS production of same-sign
$W$ pairs decaying leptonically, {\em i.e.},
\begin{equation}
    p p \to j j W^\pm W^\pm \to   jj \ell^\pm \ell^{\pm} \slashed{E}_T \;,
    \label{eq:proc}
\end{equation}
with $\ell=e,\mu$. Due to its distinctive features and low background, it is often dubbed
as the ``golden channel’'.  Previously, the ATLAS~\cite{ATLAS:2014jzl,ATLAS:2016snd,ATLAS:2019cbr,ATLAS:2023sua,ATLAS:2025wuw} and CMS~\cite{CMS:2014mra,CMS:2017fhs,CMS:2020gfh,CMS:2026gqm}
collaborations have studied this channel through VBS measurements and searches for aQGCs. In parallel, it has also been  investigated in phenomenological studies from several complementary perspectives~\cite{Fuks:2020att,Ballestrero:2020qgv,Aoki:2020til,Chaudhary:2019aim,Kalinowski:2018oxd}, while substantial progress has been made in improving the theoretical precision~\cite{Ballestrero:2018anz,Denner:2024tlu,Jager:2024eet,Dittmaier:2023nac}. Here, we analyze the $W$ boson polarization and spin correlation
asymmetries, extracted from the angular distributions of the final-state leptons, to estimate the LHC potential
to constrain  aQCGs. Our results demonstrate that spin asymmetries yield bounds comparable to those obtained from transverse invariant mass–based analyses, while their combination leads to  stronger constraints. 
\smallskip

In order to take full advantage of the spin asymmetries, we need to
know the decaying $W^\prime$s helicity frame~\cite{Leader:2001nas,Boudjema:2009fz}
to reconstruct the charged lepton in this frame. This is not a simple task due to the presence of missing neutrinos. Therefore, we adopted machine learning (ML) based regression algorithm to reconstruct the momenta of two missing neutrinos.\smallskip

In a model independent approach, we parametrize the departures from
the SM predictions to QGC using Effective Field Theory (EFT) to encode
indirect effects of BSM physics. More specifically, we assume that the
Higgs-like state observed at the LHC in 2012~\cite{Aad:2012tfa,
Chatrchyan:2012xdj} belongs to a $SU(2)_L$ doublet, allowing the
linear realization of the SM gauge symmetry in the low-energy
effective theory; this is the so called Standard Model  Effective Field
Theory (SMEFT). In this scenario, departures from the SM predictions
are parametrized by higher dimension operators as
\begin{equation}
{\cal L}_{\rm eff}  = {\cal L}_{\rm SM} + \sum_{n>4}^j
\frac{f_{n,j}}{\Lambda^{n-4}} {\cal O}_{n,j} ,
\end{equation}
where $\Lambda$ is a characteristic energy scale and ${\cal O}_{n,j}$
are higher dimensional ($n>4$) operators. The $f_{n,j}$ are the associated Wilson coefficients (WCs) which encodes the effects of BSM physics. In this framework, the lowest
dimension operators contributing to the LHC physics are of dimension
six~\cite{Degrande:2013rea}; however, they are accompanied by anomalous
triple gauge couplings that are subject to tight limits. On the other hand, genuine QGC operators are
dimension eight. 
\smallskip

The rest of the article is organized as follows: In section~\ref{sec:anaframe}, we discuss the relevant bosonic dimension-eight operators affecting the quartic $W$ vertex, along with a brief discussion on the joint density matrix  representing two $W$ bosons. The parameters of the density matrix are reconstructed as asymmetries in the angular distribution of final decayed leptons. We also discuss on reconstruction of two missing neutrinos using a neural network-based regression algorithm. In section~\ref{sec:res}, we discuss the limits on WCs using spin asymmetries and di-boson transverse mass. We 
also discuss the impact of a unitarization procedure on the attainable limits. Finally, we conclude in section~\ref{sec:conclude}.

%%%%%%%%%%%%%%%%%%%%%%%%%%%%%%%%%%%%%%%%%%%%%%%%%%%%%%%%%%%%%%%%%%%%%%
\section{Analyses framework}
\label{sec:anaframe}

\begin{figure*}[!htb] \centering
\includegraphics[width=0.32\textwidth]{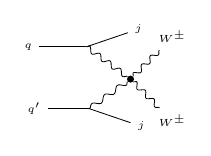}
\includegraphics[width=0.32\textwidth]{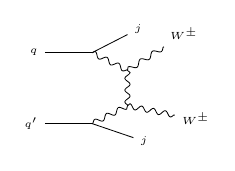}
\includegraphics[width=0.32\textwidth]{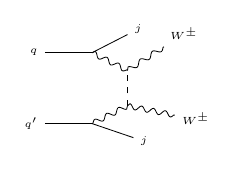}
\includegraphics[width=0.32\textwidth]{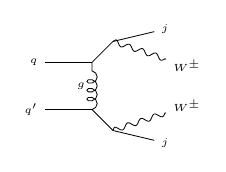}
\includegraphics[width=0.32\textwidth]{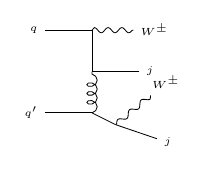}
\includegraphics[width=0.32\textwidth]{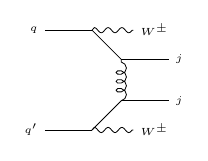}
\caption{Representative leading-order Feynman diagrams for VBS production of
same-sign $W$ pairs~($pp \to W^\pm W^\pm jj$) including pure
electroweak contribution (top row) and mixed QCD-electroweak ones
(bottom row). Anomalous quartic $W$ coupling is represented by the shaded blob in the left-top panel.}
\label{fig:feynman}
\end{figure*}

Fig.~\ref{fig:feynman} depicts some typical Feynman diagrams contributing to the
production of same-sign $W$ pairs in association with two jets. The left-top diagram represents the VBS of the same-sign $W$ bosons which receives aQGC contribution through the shaded blob.   Within the SM framework, this
process exhibits cancellations between the different contributions, preventing its cross-section growth with the center-of-mass
energy~\cite{Lee:1977eg}. Hence, VBS provides important precision
measurements that can be used to probe the gauge structure and the
electroweak symmetry breaking (EWSB)~\cite{Englert:1964et,
Higgs:1964pj, Guralnik:1964eu} sector of the SM. \smallskip

At the Born level, VBS process receive contributions from pure
electroweak interactions  (${\cal O}(\alpha^4)$), shown in the top row in Fig.~\ref{fig:feynman}, as well
as from mixed QCD-electroweak interactions
(${\cal O}(\alpha_S^2\alpha^2)$), shown in the bottom row and their small
interference term at ${\cal O}(\alpha_S\alpha^3)$.  The
same-sign $W$ boson VBS has the largest cross-section ratio of electroweak to strong
production compared to other VBS processes, as there is  no gluon
initiated processes at leading order. \smallskip

%%%%%%%%%%%%%%%%%%%%%%%%%%%%%%%%%%%%%%%%%%%%%%%%%%%%%%%%%%%%%%%%%%%%%%
%%%%%%%%%%%%%%%%%%%%%%%%%%%%%%%%%%%%%%%%%%%%%%%%%%%%%%%%%%%%%%%%%%%%%%

Focusing exclusively on anomalous quartic $WWWW$ interactions,
the C- and P-conserving dimension-eight operators that contain genuine QGC and contribute to this process are~\cite{Eboli:2006wa}
\begin{widetext}
    \begin{align}
      \mathscr{O}_{S0} &= \left[\left(D_\mu\Phi\right)^\dagger
        D_\nu\Phi\right]\times\left[\left(D^\mu
          \Phi\right)^\dagger D^\nu\Phi\right],\quad \quad
			\mathscr{O}_{S1} =
                        \left[\left(D_\mu\Phi\right)^\dagger
                          D^\mu\Phi\right]\times
                        \left[\left(D_\nu\Phi\right)^\dagger D^\nu\Phi\right], \nonumber\\
			\mathscr{O}_{S2} &=
                        \left[\left(D_\mu\Phi\right)^\dagger
                          D^\mu\Phi\right]\times\left[\left(D_\nu\Phi\right)^\dagger D^\nu\Phi\right],\quad \quad
			\mathscr{O}_{M0} =
                        \text{Tr}\left[W_{\mu\nu}W^{\mu\nu}\right]
                        \times \left[\left(D_\beta\Phi\right)^\dagger D^\beta\Phi\right], \nonumber\\
			\mathscr{O}_{M1} &= \text{Tr}\left[\widehat
                          W_{\mu\nu} \widehat
                          W^{\nu\beta}\right]\times
                        \left[\left(D_\beta\Phi\right)^\dagger
                          D^\mu\Phi\right],\quad\quad
			\mathscr{O}_{M7} =
                        \left[\left(D_\mu\Phi\right)^\dagger \widehat
                          W_{\beta\nu} \widehat W^{\beta\mu}D^\nu\Phi\right], \nonumber\\
			\mathscr{O}_{T0} &= \text{Tr}\left[\widehat
                          W_{\mu\nu} \widehat W^{\mu\nu}\right]\times
                        \text{Tr}\left[\widehat W_{\alpha\beta}
                          \widehat W^{\alpha\beta}\right],\quad \quad  
			\mathscr{O}_{T1} = \text{Tr}\left[\widehat
                          W_{\alpha\nu} \widehat
                          W^{\mu\beta}\right]\times
                        \text{Tr}\left[\widehat W_{\mu\beta} \widehat W^{\alpha\nu}\right],\nonumber\\
			\mathscr{O}_{T2} &= \text{Tr}\left[\widehat
                          W_{\alpha\mu} \widehat
                          W^{\mu\beta}\right]\times
                        \text{Tr}\left[\widehat W_{\beta\nu} \widehat W^{\nu\alpha}\right] \;,
        \end{align}
	\label{eqn:dim8op}
\end{widetext}
where, according to our conventions,  $\Phi$ stands for the Higgs doublet and its  covariant derivative is $D_\mu\Phi = \left(\partial_\mu + igW^j_\mu\frac{\tau^j}{2}+\frac{ig^\prime}{2} B_\mu \right)\Phi$
while the $W$ field strength tensor is $\widehat{W}_{\mu\nu} = \frac{i}{2}g\tau^i\left(\partial_\mu W^i_\nu -
\partial_\nu W^i_\mu + g\epsilon_{ijk} W^j_\mu W^k_\nu \right)$.  We
denote the Pauli matrices as $\tau^j$.  The operators $\mathscr{O}_{S1}$ and $\mathscr{O}_{S2}$ contain the same $WWWW$
vertex in the same-sign VBS~\cite{Eboli:2016kko}, therefore, we  consider only
$\mathscr{O}_{S1}$ in the current analyses. \smallskip

In the current article, we exploit the information of the spin of the gauge boson to construct the related polarization and spin correlation parameters with the objective of constraining aQGC. Along with the spin related observables, we also compute the one-dimensional distribution in transverse mass of the $WW$ state and use it as well to probe the anomalous couplings. \smallskip

The quantum state of the $W$ boson can be represented by a density matrix ($\rho_W$) of
size $3\times 3$  with eight independent parameters that, in Cartesian form, can be written as~\cite{Bourrely:1980mr}
\begin{equation}
\rho_W(\lambda_W,\lambda_W^\prime) =
\frac{1}{3}\left[\mathbb{I}+\frac{3}{2}p_i\cdot S_i
+ \sqrt{\frac{3}{2}}T_{ij}\{S_i,S_j\}\right],
\end{equation}
where  $S_i$ are three spin-1 fundamental operators and $\{S_i, S_j\}$ stands for their
anti-commutation.  Moreover,  $p_i$ are three vector polarizations and $T_{ij}$ is the
traceless polarization tensor of $W$ boson. 
These eight polarizations can be obtained from the
angular distribution of final decayed leptons in the rest frame of $W$
boson that is given by~\cite{Boudjema:2009fz} 
\begin{widetext}
\begin{align}
\label{eq:diffone}
\frac{1}{\sigma}\frac{d\sigma}{d\Omega_\ell} &=
\frac{3}{8\pi}\left[\left(\frac{2}{3}-(1-3\delta)\frac{T_{zz}}{\sqrt{6}}\right)
+ \alpha p_z \cos\theta_\ell +
\sqrt{\frac{3}{2}}(1-3\delta)T_{zz}\cos^2\theta_\ell \right.\nonumber\\&+
\left. \left(\alpha p_x +
2\sqrt{\frac{2}{3}}(1-3\delta)T_{xz}\cos\theta_\ell\right)\sin\theta_\ell
\cos\phi_\ell + \left(\alpha p_y
+2\sqrt{\frac{2}{3}}(1-3\delta)T_{yz}\cos\theta_\ell\right)\sin\theta_\ell
\sin\phi_\ell \right.\nonumber\\&+
\left. (1-3\delta)\left(\frac{T_{x^2-y^2}}{\sqrt{6}}\cos(2\phi_\ell)
+ \sqrt{\frac{2}{3}}T_{xy}\sin(2\phi_\ell)\right)\sin^2\theta_\ell
\right]\;,
\end{align}
\end{widetext}
where $\theta_\ell$ and $\phi_\ell$ are the polar and azimuthal angles of the
final lepton in the rest frame of the $W$ boson with its would-be momentum along the $z$-axis. 
The spin analysis
parameter is $\alpha=-1$ owing to chiral nature of $W$ coupling to
leptons and $m_W \gg m_\ell$. Assuming the lepton to be massless in the
large energy limit, the parameter $\delta$ vanishes~\cite{Boudjema:2009fz}. 
\smallskip

Polarization parameters of $W$ boson can  be obtained from  asymmetries
associated with angular functions of final leptons using the angular distribution given in Eq.~(\ref{eq:diffone}). For example, $p_x$
and $T_{yz}$ can be obtained as~\cite{Rahaman:2021fcz}
\begin{align}
\mathcal{A}_x &= \frac{\sigma(\sin\theta_\ell\cos\phi_\ell > 0)-\sigma(\sin\theta_\ell\cos\phi_\ell < 0)}{\sigma(\sin\theta_\ell\cos\phi_\ell > 0)+\sigma(\sin\theta_\ell\cos\phi_\ell < 0)} \nonumber\\&= \frac{3}{4}\alpha p_x\;,\nonumber\\
\mathcal{A}_{yz} &=\frac{\sigma(\sin\theta_\ell\cos\theta_\ell\sin\phi_\ell > 0)-\sigma(\sin\theta_\ell\cos\theta_\ell\sin\phi_\ell < 0)}{\sigma(\sin\theta_\ell\sin\theta_\ell\sin\phi_\ell > 0)+\sigma(\sin\theta_\ell\sin\theta_\ell\sin\phi_\ell < 0)} \nonumber\\
&= \frac{2}{\pi}\sqrt{\frac{2}{3}}(1-3\delta)T_{yz}\;.
\end{align}
In general,  asymmetries are defined by
\begin{align}
\mathcal{A}_i = \frac{\sigma(c_i > 0) - \sigma(c_i < 0)}{\sigma(c_i >0) + \sigma(c_i < 0)},~ i\in {1,2,\dots,8}\;,
\end{align}
where the correlators ($c_i$) are listed in Table~\ref{tab:correlators}.

In the case when two $W$ bosons are co-produced, we can represent the $WW$ quantum system by the joint density matrix~\cite{Rahaman:2021fcz}
\begin{widetext}
\begin{align}
\label{eq:rho11}
&\rho(\lambda_{W_1},\lambda_{W_2},\lambda_{W_1}^\prime,\lambda_{W_2}^\prime)
= \frac{1}{9}\left[\mathbb{I}+\frac{3}{2}p_i^{W_1}\cdot S_i
\otimes \mathbb{I} + \frac{3}{2}\mathbb{I}\otimes
p_i^{W_2}\cdot S_i + \frac{3}{2}T^{W_1}_{ij}\{S_i,S_j\}\otimes
\mathbb{I}+\frac{3}{2}\mathbb{I}\otimes
T_{ij}^{W_2}\{S_i,S_j\}+\right.\nonumber\\&\left. \frac{9}{4}pp^{W_1W_2}_{ij}S_i\otimes
S_j + \frac{3}{2}\sqrt{\frac{3}{2}}pT^{W_1W_2}_{ijk}S_i
\otimes \{S_j,S_k\} +
\frac{3}{2}\sqrt{\frac{3}{2}}Tp_{ijk}^{W_1W_2}\{S_i,S_j\}\otimes
S_k + \frac{3}{2}TT^{W_1W_2}_{ijkl}\{S_i,S_j\}\otimes
\{S_k,S_l\}\right] \;.
\end{align}
\end{widetext}
Here, $pp^{W_1W_2}$, $pT^{W_1W_2}=Tp^{W_2W_1}$ and $TT^{W_1W_2}$ represent
vector-vector, vector-tensor and tensor-tensor spin correlations among
two $W$ bosons, respectively. These additional $64$ spin-correlation
parameters can also be obtained from the joint angular distribution of the two
decayed final leptons in the rest frame of their respective mother particle. \smallskip
%

%%%%%%%%%%%%%%%%%%%%%%%%%%%%%%%%%%%%%%%%%%%%%%
\begin{table}[!htb]
\centering
\caption{\label{tab:correlators}List of asymmetries related to the eight polarization parameters of $W$ boson and the associated correlators or angular functions.}
\renewcommand{\arraystretch}{1.5}
\begin{tabular*}{0.45\textwidth}{@{\extracolsep{\fill}}lcl@{}}\hline
Asymmetries & Correlators & Angular Functions\\
\hline
$A_1 \equiv A_x $ & $c_1$ & $\sin\theta_\ell\cos\phi_\ell$\\
$A_2 \equiv A_y $ & $c_2$ & $\sin\theta_\ell\sin\phi_\ell$\\
$A_3 \equiv A_z $ & $c_3$ & $\cos\theta_\ell$\\
$A_4 \equiv A_{xy} $ & $c_4$ & $\sin^2\theta_\ell\sin(2\phi_\ell)$\\
$A_5 \equiv A_{xz} $ & $c_5$ & $\sin\theta_\ell\cos\theta_\ell\cos\phi_\ell$\\
$A_6 \equiv A_{yz} $ & $c_6$ & $\sin\theta_\ell\cos\theta_\ell\sin\phi_\ell$\\
$A_7 \equiv A_{x^2-y^2} $ & $c_7$ & $\sin^2\theta_\ell\cos(2\phi_\ell)$\\
$A_8 \equiv A_{zz} $ & $c_8$ & $\sin(3\theta_\ell)$\\
\hline
\end{tabular*}
\end{table}
%%%%%%%%%%%%%%%%%%%%%%%%%%%%%%%%%%%%%%%%%%%%%%%%

In the case of  pair production of  $W$ bosons, the normalized joint angular distribution of the two final state leptons is given by~\cite{Rahaman:2021fcz}
\begin{align}
\label{eq:diffrate}
\frac{1}{\sigma}\frac{d^2\sigma}{d\Omega^{\ell_1}d\Omega^{\ell_2}} &= \left(\frac{3}{4\pi}\right)^2\sum_{\mathrm{All}~\lambda}\rho(\lambda_1,\lambda_1^\prime,\lambda_2,\lambda_2^\prime) \nonumber\\&\otimes  \Gamma_{W^1}\left(\lambda_1,\lambda_1^\prime\right)\otimes \Gamma_{W^2}\left(\lambda_2,\lambda_2^\prime\right) \;,
\end{align}
where $W^{1}$ and $W^2$ represent the $W$ boson associated with the hardest and second hardest leptons, respectively. Here, $\lambda \in \left[-1,0,+1\right]$ are the helicities of the $W$ bosons and $\Gamma$ is its  decay density matrix~\cite{Boudjema:2009fz}. \smallskip

The 16 polarization and 64 spin correlation parameters of the joint density matrix can be obtained from asymmetries as in the case of a single $W$ boson. For example, correlation among vector component of two $W$ bosons, denoted as $pp_{zz}^{WW}$ in Eq.~\eqref{eq:rho11} is obtained as~\cite{Rahaman:2021fcz}
\begin{align}
\mathcal{A}[pp^{W_1W_2}_{zz}] &=  \frac{\sigma( \cos\theta_{\ell_1}\cos\theta_{\ell_2} >0 )-\sigma( \cos\theta_{\ell_1}\cos\theta_{\ell_2} <0 )}{\sigma( \cos\theta_{\ell_1}\cos\theta_{\ell_2} >0 )+\sigma( \cos\theta_{\ell_1}\cos\theta_{\ell_2} <0 )}
\notag\\&= \frac{1}{4}\alpha^2 pp_{zz}^{W_1W_2} \;.
\end{align}
%
%\begin{align}
%\mathcal{A}[pp^{W_1W_2}_{ij}] &= \int d\Omega_{\ell_1}\int d\Omega_{\ell_2} \left(\frac{1}{\sigma}\frac{d^2\sigma}{d\Omega_{\ell_1}d\Omega_{\ell_2}}\right)\nonumber\\&= \frac{1}{4}\alpha^2 pp_{ij}^{W_1W_2} \;. 
%\end{align}
%
In general, the expression of spin correlation asymmetries for numerical calculation is given by~\cite{Rahaman:2021fcz}
\begin{equation}
\mathcal{A}_{ij} = \frac{\sigma(c_i^{\ell_1}c_j^{\ell_2} > 0) - \sigma(c_i^{\ell_1}c_j^{\ell_2} < 0)}{\sigma(c_i^{\ell_1}c_j^{\ell_2} > 0) + \sigma(c_i^{\ell_1}c_j^{\ell_2} < 0)},~i,j \in {1,\cdots,8},
\end{equation}
where the correlators are listed in Table~\ref{tab:correlators}. For details on the relation between the asymmetries and the parameters of the density matrix given in Eq.~\eqref{eq:rho11}, see Ref.~\cite{Rahaman:2021fcz}.\smallskip

The reconstruction of the joint density matrix at the decay level is  non-trivial due to the presence of  neutrinos in the final state. Ergo, an additional neutrino reconstruction scheme needs to be employed to obtain the neutrino momenta. We use a multi-layer perceptron (MLP) method to reconstruct neutrino momenta.\smallskip

The $jj \ell^\pm \ell^{\pm} \slashed{E}_T$ final state receives non-resonant contributions in addition to the resonant process given in Eq.~(\ref{eq:proc})~\cite{Ballestrero:2018anz}. To account for all such contributions, we simulate the full process
\begin{equation}\label{eq:proc-mg5}
p p \to  jj \ell^\pm \ell^{\pm} \slashed{E}_T
\end{equation}
using \textsc{MadGraph5\_aMC@NLO}~\cite{Alwall:2014hca}  at the leading order  in QCD for an LHC  center-of-mass energy   of  13.6 TeV.  In order to mitigate the irreducible  QCD contributions, we apply the VBS selection cuts adopted in the CMS study~\cite{CMS:2020gfh}:
\begin{align}
\label{eqn:vbscuts}
&m_{jj} > 500~\text{GeV}, \quad p_{T}^{j} > 50~\text{GeV},
\quad \slashed{p}_T > 30~\text{GeV},\nonumber\\ &p_T^{\ell_1} >
25~\text{GeV},\quad p_T^{\ell_2} > 20~\text{GeV}, \quad m_{\ell\ell} >
20~\text{GeV},\nonumber\\ &|\Delta\eta_{jj}| > 2.5,\quad
\text{max}(z_\ell^\star) < 0.75  \;.
\end{align}
Here, the two charged leptons are ordered according to transverse momenta ($p_T$), denoting the hardest one
by $\ell_1$ and the other by $\ell_2$. The leptonic Zeppenfield variable ($z_\ell^\star$)  is defined in terms of the pseudorapidites of the two hardest jets and the lepton one as
\begin{equation}
z_\ell^\star = \left |\eta^\ell
-(\eta^{j_1}+\eta^{j_2})/2 \right|/\left|\Delta\eta_{jj} \right|.
\end{equation}
%}

In our analyses,  leptons are considered  isolated if the  activity around a cone size of  $\Delta R_{\rm Max}=0.4$ is
sufficiently small, corresponding to at most 12\% (25\%) of the lepton transverse momentum for electrons (muons). The final jets are reconstructed using the \textsc{FastJet} package~\cite{Cacciari:2011ma}  with \textsc{anti-$k_T$} clustering algorithm with jet radius of $R=0.4$.  Parton shower and hadronization of the generated events were performed using  \textsc{Pythia}~\cite{Sjostrand:2007gs} while the fast detector simulation was carried out  with  \textsc{Delphes}~\cite{deFavereau:2013fsa} using the default CMS card. In addition, events are  filtered  using the  $b$-jet tagging algorithm available in \textsc{Delphes}. Finally,  we select events containing two same-sign leptons, two jets, along with missing energy after the VBS selection cuts given in Eq.~(\ref{eqn:vbscuts}). \smallskip

%%%%%%%%%%%%%%%%%%%%%%%%%%%%%%%%%%%%%%%%%%%%%%%%%%%%%%%
\begin{figure*}[!htb]
\centering
\includegraphics[width=0.325\textwidth]{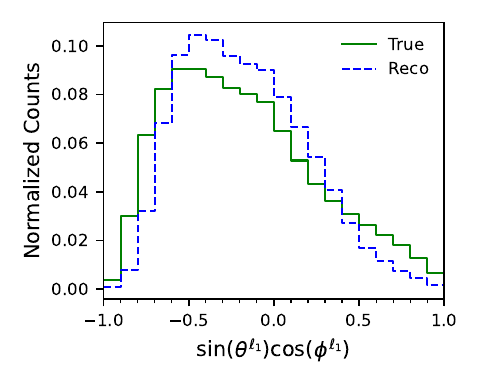}
\includegraphics[width=0.325\textwidth]{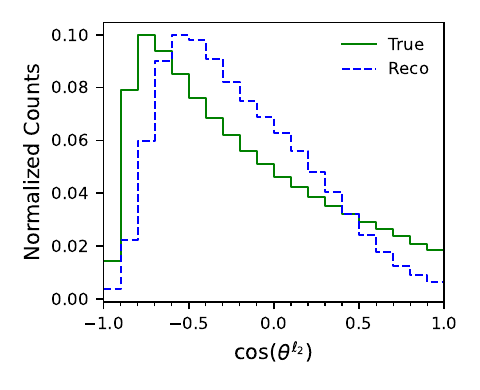}
\includegraphics[width=0.325\textwidth]{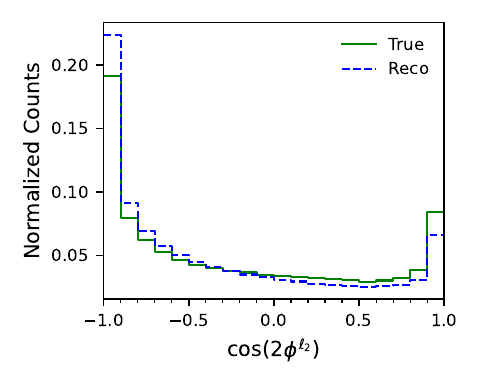}
\includegraphics[width=0.325\textwidth]{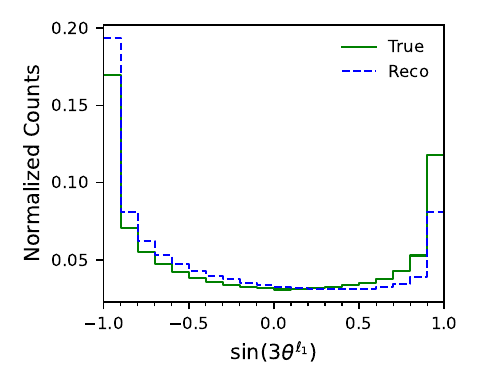}
\includegraphics[width=0.325\textwidth]{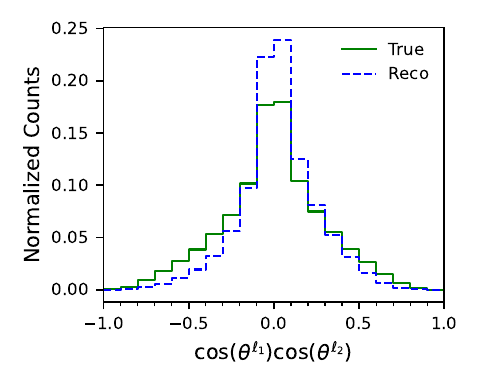}
\includegraphics[width=0.325\textwidth]{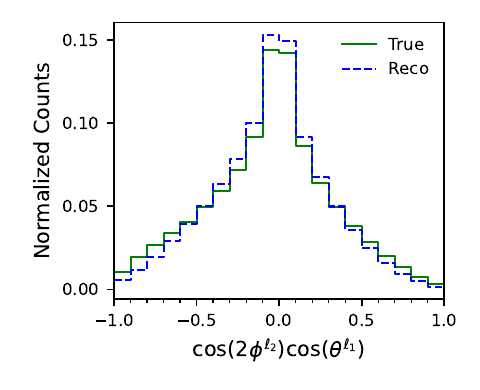}
\caption{\label{fig:phicost} Normalized parton level distributions for leptonic angular variables  related to the polarization and spin correlations of the $W^\pm W^\pm$ system (see Table~\ref{tab:correlators}) using the truth-level information of the neutrinos (labeled True) and reconstructed momenta of neutrinos using MLP (labeled Reco). Here, we applied the cuts in Eq.~(\ref{eqn:vbscuts}) and  $\phi$ ($\theta$) stands for the azimuthal (polar) angle in the helicity frame~\cite{Leader:2001nas,Boudjema:2009fz}.   }
\end{figure*}
%%%%%%%%%%%%%%%%%%%%%%%%%%%%%%%%%%%%%%%%%%%%%%%%%%%%%%%

To construct the polarization and spin correlations in the rest frame of the $W$ bosons, individual momenta of neutrinos are needed.  We took advantage of the predictive power of a MLP  to perform a six-variable regression analysis in order to reconstruct the neutrino's momenta.  The four momenta, transverse momenta ($p_T$), azimuthal orientation ($\phi$) and  pseudorapidity ($\eta$) of the two jets and two leptons, along with the missing transverse energy, are  used in the MLP network:
\begin{itemize}
\item Jets: $p_x^j,~p_y^j,~p_z^j,~E_j,~p_T^j,~\eta^j~\phi^j.$
\item Lepton: $p_x^{\ell},~p_y^{\ell},~p_z^{\ell},~E^{\ell},~p_T^{\ell},~\eta^{\ell},~\phi^{\ell}~.$
\item MET: $\slashed{E}_T$,~$\phi^{\text{Miss}}$.
\end{itemize}
Here, $\slashed{E}_T$ is the missing transverse momenta and $\phi^{\mathrm{Miss}} = \tan^{-1}(E_y^{\mathrm{Miss}}/E_x^{\mathrm{Miss}})$~\cite{ATLAS:2018txj} denotes the azimuthal orientation of the $\slashed{E}_T$ vector in the transverse plane. 
These input features are then fed into a MLP implemented using \textsc{TensorFlow}~\cite{Abadi:2016kic} with  four hidden layers containing 200, 100, 50, and 25 nodes, respectively. Input features are standardized using \textsc{StandardScaler} from \textsc{scikit-learn}~\cite{Pedregosa:2011ork}. The network uses \textsc{ReLU} activation function, mean squared error as the loss function with \textsc{L2} regularization, and is optimized using the \textsc{Adam} optimizer.
The parton level generated SM events are used to train and test the MLP network after the VBS selection cuts in Eq.~(\ref{eqn:vbscuts}).
Out of $2\times 10^6$ simulated events, $70\%$ are used for training, while the remaining $30\%$ are equally split between validation and testing the network. 
\smallskip

In our analyses, we used the reconstructed neutrino momenta to perform
the Lorentz transformations to the $W$'s helicity frame. In order to
probe the quality of the reconstructed angular distributions of the
charged leptons, Fig.~\ref{fig:phicost} exhibits the normalized
angular spectra  where we show the true distributions as well as
the reconstructed ones by MLP.
For demonstration purposes, we present a set of basic angular functions (see Table~\ref{tab:correlators}) for polarization, namely $\sin\theta\cos\phi$, $\cos\theta$, $\cos(2\phi)$, and $\sin(3\theta)$, and for spin correlations, $\cos\theta^{\ell_1}\cos\theta^{\ell_2}$ and $\cos(2\phi^{\ell_2})\cos\theta^{\ell_1}$. All angular functions show reasonable reconstruction, except for $\cos\theta^{\ell_2}$, which exhibits a shift in the peak. \smallskip

In our analyses, we use the detector level angular asymmetries related to polarization and spin correlations and the distribution of  $WW$ transverse mass 
%($m_T^{WW}$) 
defined as
\begin{equation}
	m_T^{WW} = \sqrt{m_{\ell\ell}^2 + 2(E_{T}^{\ell\ell}E_{T}^{\rm Miss} - \vec{p}_{T}^{\,\ell\ell}\cdot \vec{p}_{T}^{\rm \, Miss})} \;,
\end{equation}
to obtain the constraints on aQGCs. We consider the SM and aQGC contributions as well as the reducible backgrounds coming from the VBS production of $WZ$ and $tZj$.  \smallskip

%%%%%%%%%%%%%%%%%%%%%%%%%%%%%%%%%%%%%%%%%%%%%%%%%%%%%%%%%%%%%%%%%%%%%%

%%%%%%%%%%%%%%%%%%%%%%%%%%%%%%%%%%%%%%%%%%%%%%%%%%%%%%%
\section{Results}
\label{sec:res}
%%%%%%%%
\begin{figure*}[!htb]
\centering
\includegraphics[width=0.4\textwidth]{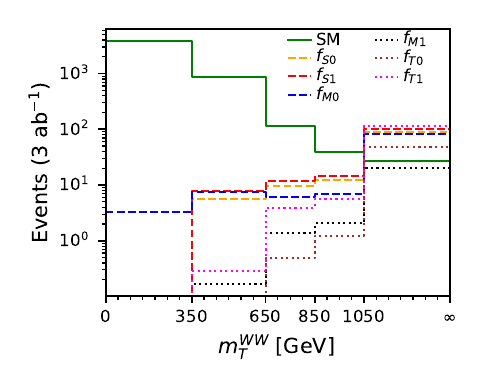}
\includegraphics[width=0.4\textwidth]{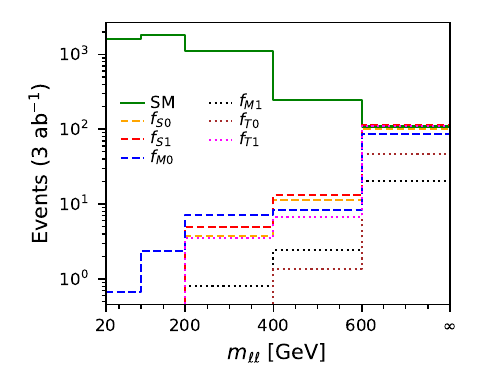}
\includegraphics[width=0.4\textwidth]{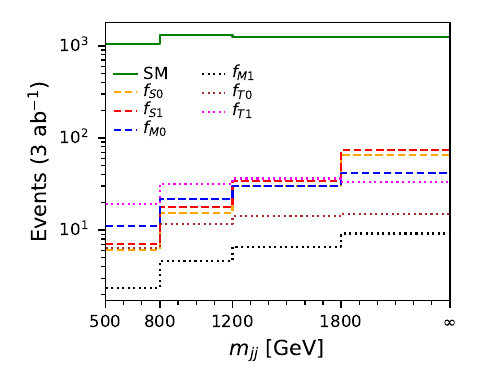}
\includegraphics[width=0.4\textwidth]{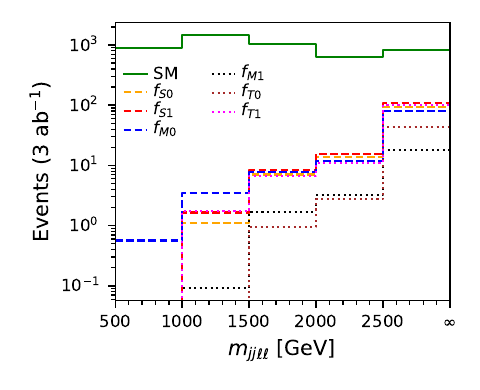}
\caption{\label{fig:kinematics} aQGC-sensitive kinematic distributions at detector level for an integrated luminosity of 3 ab$^{-1}$ luminosity. We considered  one non-vanishing WC one at a time with their values given in Eq.~(\ref{eq:bp}).}
\end{figure*}
%%%%%%%%%%%%%%%%%%%%%%%%%%%%%%%%%%%%%%%%%%%%%%%%%%%%%%%
\begin{figure*}[!htb]
\centering
\includegraphics[width=0.325\textwidth]{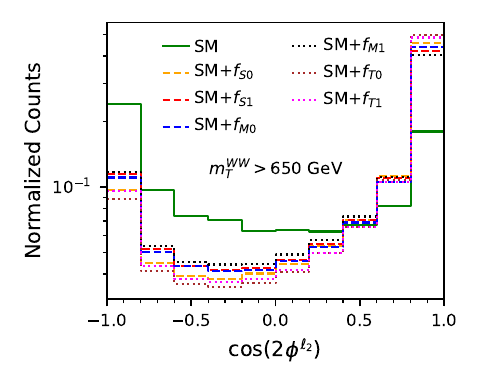}
\includegraphics[width=0.325\textwidth]{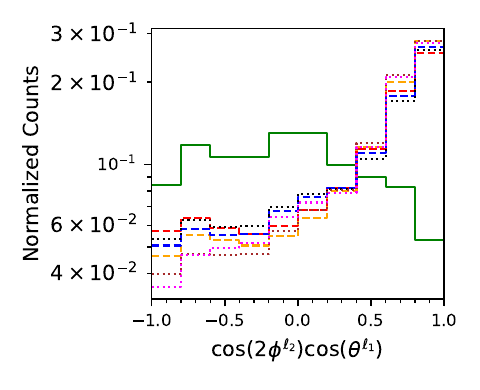}
\includegraphics[width=0.325\textwidth]{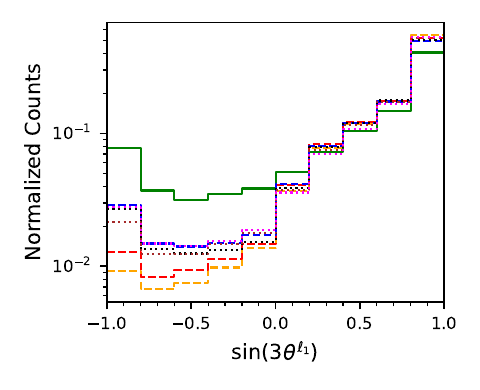}
\includegraphics[width=0.325\textwidth]{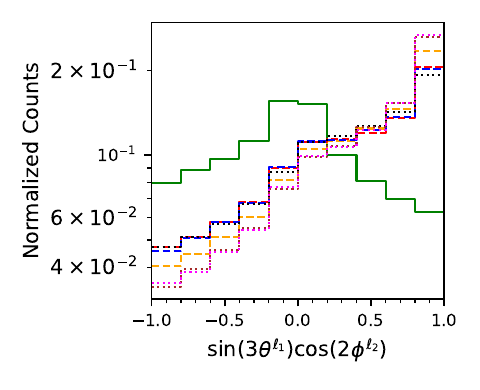}
\includegraphics[width=0.325\textwidth]{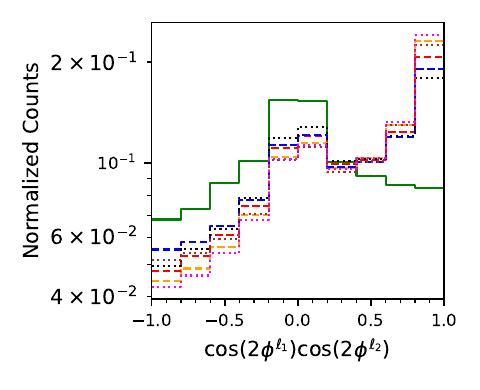}
\includegraphics[width=0.325\textwidth]{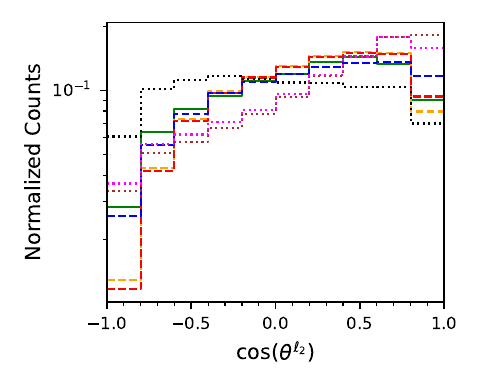}
\caption{\label{fig:recoangcomapre}  aQGC-sensitive normalized angular distributions 
 related to the polarization and spin correlations of the $WW$ system for $m_T^{WW}>650$ GeV.  The distributions  are obtained by adding the SM contribution and switching on one WC at a time using the benchmark values given in Eq.~(\ref{eq:bp}).
}
\end{figure*}
%%%%%%%%%%%%%%%%%%%%%%%%%%%%%%%%%%%%%%%%%%%%%%%%%%%%%%%

In Table~\ref{tab:bkg-cs-events}, we list the cross sections after VBS cuts including leptonic decays, detector efficiencies  and the corresponding number of events for integrated luminosities of 137 and 3000~fb$^{-1}$ for the SM same-sign VBS ($WWjj$)  as well as the $WZjj$ and $tZj$ backgrounds.
The number of events for an integrated luminosity of 137~fb$^{-1}$ is shown for comparison with the expected yields reported in the CMS study~\cite{CMS:2020gfh}. To account for additional background contributions such as non-prompt and wrong-sign leptons, which are included in the CMS analysis but not explicitly simulated in our study, we rescale our total SM background (by a factor $\sim 2$) to the total background in the CMS study. \smallskip

%%%%%%%%%%%%%%%%%%%%%%%%%%%%%%%%%%%%%%%%%%%%%%%%
\begin{table}[!h]
\caption{Cross sections ($\sigma$) including leptonic decays, detector efficiencies ($\epsilon$) and number of events for luminosities of 137 and 3000 fb$^{-1}$ for the SM same-sign VBS ($jj\ell^\pm\ell^\pm\slashed{E}_T$) and $WZjj$ and $tZj$ backgrounds after the VBS cuts in Eq.~(\ref{eqn:vbscuts}).}
\label{tab:bkg-cs-events}
\renewcommand{\arraystretch}{1.5}
\centering
\begin{tabular*}{0.48\textwidth}{@{\extracolsep{\fill}}lcccc@{}}\hline
&&&\multicolumn{2}{c}{Number of events} \\ \hline
Process & $\sigma$ (fb) & $\epsilon$ (\%)  & 137 fb$^{-1}$  & 3000 fb$^{-1}$ \\\hline
$jj\ell^\pm\ell^\pm\slashed{E}_T$ & 6.22 & 26  & 221.6 & 4852 \\
$WZjj$ &12.4 & 2.4  & 40.7 & 893 \\
$tZj+1j$& 7.86 & 0.59  & 6.4 & 139 \\ \hline
\end{tabular*}
\end{table}
%%%%%%%%%%%%%%%%%%%%%%%%%%%%%%%%%%%%%%%%%%%%%%%%%%%%%%

We begin our analyses by studying several kinematic distributions,  such as transverse and invariant masses and leptonic angular distributions  related to the polarization and spin correlations  after reconstructing the neutrinos using MLP as detailed in the previous section. We analyze the distributions for the  SM as well as for eight aQGC benchmark points (BPs) chosen based on the CMS study~\cite{CMS:2020gfh}, with only one WC being varied at a time to isolate its impact. The chosen BPs are
\begin{eqnarray}\label{eq:bp}
&&f_{S0} = 6.0,~  f_{S1} = 19.0,~\notag\\
&&f_{M0} = 3.0,~ f_{M1} = 4.7, ~ f_{M7} = 7.0,~\notag\\
&&f_{T0} = 0.2,~  f_{T1} = 0.15,~ f_{T2} = 0.40,
\end{eqnarray}
in TeV$^{-4}$ units. \smallskip

Fig.~\ref{fig:kinematics} depicts  the distributions of the transverse mass of the $WW$ system, 
%($m_T^{WW}$),  
the invariant mass of the two leptons ($m_{\ell\ell}$), the invariant mass of the dijet system ($m_{jj}$), and the invariant mass of the jets and leptons ($m_{jj\ell\ell}$)  for an integrated luminosity of 3~ab$^{-1}$. The aQGC curves show only the anomalous contribution arising from a single non-vanishing aQGC at a time. The two distributions shown in the top panel exhibit good sensitivity to aQGCs, as the SM contribution falls off in the high-energy tails while the aQGC contributions increase.  Notice that  $m_T^{WW}$ presents a stronger sensitivity to aQGCs than $m_{\ell\ell}$. In contrast, the two distributions in the bottom panel are less sensitive to aQGCs, since the SM contribution remains relatively flat. In our analyses, we therefore use the $m_T^{WW}$ distribution, together with angular asymmetries, to constrain the aQGCs.  \smallskip

We analyze the angular distributions in the regions $350~\mathrm{GeV} \leq m_T^{WW} < 650~\mathrm{GeV}$ and $m_T^{WW} \geq 650~\mathrm{GeV}$, and observe that the higher-$m_T^{WW}$ region is significantly more  sensitivity to the aQGCs, as expected. In Fig.~\ref{fig:recoangcomapre}, we present a set of normalized angular distributions sensitive to aQGCs, related to the polarization and spin correlations of the $WW$ system, for $m_T^{WW} > 650~\mathrm{GeV}$. The aQGC curves  contain the SM contribution  as well as  only  one anomalous coupling  at a time. We find that not only  polarization-related  spectra, such as $\cos(2\phi)$, $\sin(3\theta)$, and $\cos(\theta)$, but also spin-correlation observables, including $\cos(2\phi^{\ell_2})\cos(\theta^{\ell_1})$,  $\sin(3\theta^{\ell_1})\cos(2\phi^{\ell_2})$, and $\cos(2\phi^{\ell_1})\cos(2\phi^{\ell_2})$ exhibit strong sensitivity to aQGC effects.  In our analysis, we construct the angular asymmetries in both $m_T^{WW}$ regions to  constrain the aQGCs. \smallskip

%%%%%%%%%%%%%%%%%%%%%%%%%%%%%%%%%%%%%%%%%%%%%%%%%%%%%%%
\begin{figure*}[!htb]
\centering
\includegraphics[width=1\textwidth]{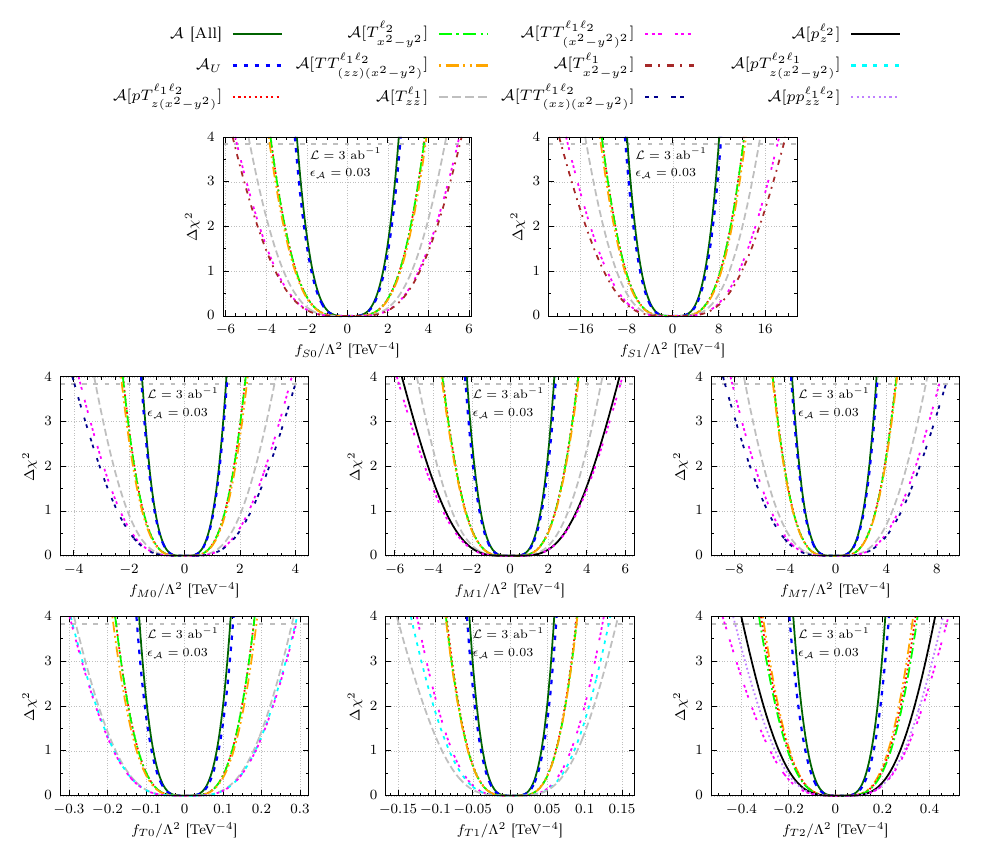}
\caption{\label{fig:sensitivity}
One-parameter $\Delta \chi^2$ as a function of the dimension-8 WCs for the six most sensitive asymmetries for each of the WCs.
We considered  an integrated luminosity of $\mathcal{L} =3$ ab$^{-1}$ and a systematic uncertainty of 0.03 on the asymmetries.
The $\Delta \chi^2$ combining  all  asymmetries  and the same with the best 10 sensitive asymmetries $(\mathcal{A}_U)$ are also shown.
}
\end{figure*}
%%%%%%%%%%%%%%%%%%%%%%%%%%%%%%%%%%%%%%%%%%%%%%%%%%%%%%%

In order to assess the potential of the above polarization and spin correlation asymmetries  to probe aQGCs, we parametrized the total production cross section as 
\begin{equation}
\label{eq:cpevenfit}
\sigma(\mathbf{f}) = \sigma_{\rm SM} + \sum_{i}f_i \sigma_i + \sum_{i}f_i^2 \sigma_{ii} +   \sum_{i\neq j}f_if_j \sigma_{ij} \;,
\end{equation}
with the WC vector
$$
\mathbf{f}=\left\{f_{S0}  ,  f_{S1}   , f_{M0}  , f_{M1}  ,  f_{M7} , f_{T0}  ,  f_{T1}  , f_{T2} \right\}.
$$
Here, $\sigma_{\rm SM}$ is the SM cross section, $\sigma_i$ stands for the interference between the SM and  aQGCs,  $\sigma_{ii}$ is  the quadratic contribution of aQGC, and $\sigma_{ij}$ represents the interference between two distinct aQGC amplitudes. For instance, Table~\ref{tab:cs-wc} contains the values of the $\sigma$'s after applying  the VBS cuts in Eq.~(\ref{eqn:vbscuts}). Since all the WCs considered in this analysis are $CP$-even, we use the parametrization given in Eq.~(\ref{eq:cpevenfit}) for the numerators of the 44 $CP$-even asymmetries\footnote{Each $W$ boson has eight polarization states, of which five are $CP$-even. Therefore, out of the total $8\times 8 = 64$ possible spin-correlation terms, the $5\times 5 = 25$ even-even and $3\times 3 = 9$ odd-odd combinations together give 34 $CP$-even spin-correlation parameters. Combined with the 10 single-particle $CP$-even polarization observables, this yields a total of 44 $CP$-even asymmetries.}, after evaluating the asymmetries up to linear and quadratic order in each aQGC.
\smallskip

%%%%%%%%%%%%%%%%%%%%%%%%%%%%%%%%%%%%%%%%%%%%%%%%%%%%%%%%%%%%%%%%%%%%%%%%%%%%%%%%%
\begin{table*}[!htb]
\caption{Coefficients for the linear, quadratic, and interference contributions by  aQGCs  for the process in Eq.~(\ref{eq:proc-mg5}) with the VBS  cuts on the jets and leptons given in Eq.~\eqref{eqn:vbscuts}.}
\label{tab:cs-wc}
\renewcommand{\arraystretch}{1.5}
\centering
\begin{tabular*}{1\textwidth}{@{\extracolsep{\fill}}lccccc@{}}\hline
WC       & linear ($\sigma_i$ fb-TeV$^4$) & quadratic ($\sigma_{ii}$ fb-TeV$^8$)  & WC-$i$ & WC-$j$ & Interference   ($\sigma_{ij}$ fb-TeV$^8$)   \\ \hline
$f_{S0}$ & $-1.74\times 10^{-3} $&$ 3.12\times 10^{-3}$ &$f_{S0}$ & $f_{S1}$ &$ 2.06\times 10^{-3}  $\\
$f_{S1}$ & $-5.39\times 10^{-4} $&$ 3.56\times 10^{-4}$ &$f_{M0}$ & $f_{M1}$ &$ 1.26\times 10^{-3}  $\\
$f_{M0}$ & $3.54\times 10^{-3}  $&$ 1.07\times 10^{-2}$ &$f_{M0}$ & $f_{M7}$ &$ 6.36\times 10^{-3}  $\\
$f_{M1}$ & $-1.12\times 10^{-3} $&$ 1.17\times 10^{-3}$ &$f_{M1}$ & $f_{M7}$ &$ -5.87\times 10^{-3} $\\
$f_{M7}$ & $2.08\times 10^{-3}  $&$ 2.17\times 10^{-3}$ &$f_{T0}$ & $f_{T1}$ &$ 5.65                $\\
$f_{T0}$ & $-4.54\times 10^{-2} $&$ 1.78              $ &$f_{T0}$ & $f_{T2}$ &$ 1.11                $\\
$f_{T1}$ & $-2.29\times 10^{-1} $&$ 8.16              $ &$f_{T1}$ & $f_{T2}$ &$ 4.34                $\\
$f_{T2}$ & $-1.15\times 10^{-1} $&$ 8.91\times 10^{-1}$ &                       &    \\\hline
\end{tabular*}
\end{table*}
%%%%%%%%%%%%%%%%%%%%%%%%%%%%%%%%%%%%%%%%%%%%%%%%%%%%%%%%%%%%%%%%%%%%%%%%%%%%%%%%%

We quantify the sensitivity of the asymmetries to the WCs in terms of a
$\Delta\chi^2$ function for each asymmetry defined by:
\begin{equation}
\label{eq:chisq}
\Delta\chi^2(\mathcal{A}_i,\mathbf{f}) =
\left(\frac{\mathcal{A}_i({\rm BKG}+\mathbf{f})-\mathcal{A}_i({\rm BKG})}{\delta\mathcal{A}_i}
\right)^2 ,
\end{equation}
with the estimated error for asymmetries given by
\begin{equation}
\delta\mathcal{A} =
\sqrt{\frac{1- \mathcal{A}^2({\rm BKG})}{\mathcal{L}~\sigma({\rm BKG})}+\epsilon_{\cal A}^2} \;,
\end{equation}
where $\epsilon_{\cal A}$ denotes the systematic uncertainty in $\mathcal{A}$,
$\mathcal{L}$ is the integrated luminosity, and ${\rm BKG}$ corresponds to the sum of all background processes considered in the analysis. \smallskip

In Fig.~\ref{fig:sensitivity}, we present the one-parameter $\Delta\chi^2$ distributions as functions of the WCs for the six most sensitive asymmetries for each WC assuming an integrated luminosity of ${\cal L}=3$~ab$^{-1}$ and a systematic uncertainty of $\epsilon_{\cal A}=0.03$. The horizontal line at $\Delta\chi^2 = 3.84$ corresponds to the 95\% confidence level (C.L.) limit on the WCs. The asymmetries $\mathcal{A}[pT_{z(x^2-y^2)}^{\ell_1\ell_2}]$, $\mathcal{A}[T_{x^2-y^2}^{\ell_2}]$, and $\mathcal{A}[TT_{(zz)(x^2-y^2)}^{\ell_1\ell_2}]$ provide comparable and among the best sensitivities to all eight aQGCs. The remaining asymmetries from the union of the best six include $\mathcal{A}[T_{zz}^{\ell_1}]$, $\mathcal{A}[TT_{(x^2-y^2)^2}^{\ell_1\ell_2}]$, $\mathcal{A}[T_{x^2-y^2}^{\ell_1}]$, $\mathcal{A}[TT_{(xz)(x^2-y^2)}^{\ell_1\ell_2}]$, $\mathcal{A}[p_z^{\ell_2}]$, $\mathcal{A}[pT_{z(x^2-y^2)}^{\ell_2\ell_1}]$, and $\mathcal{A}[pp_{zz}^{\ell_1\ell_2}]$. Overall, the spin-correlation asymmetries exhibit slightly better sensitivity than the polarization asymmetries. \smallskip

%%%%%%%%%%%%%%%%%%%%%%%%%%%%%%%%%%%%%%%%%%%%%%%%%%%%%%%%%%%%%%%%%%%%
\begin{figure*}[!htb]
\centering
\includegraphics[width=1\textwidth]{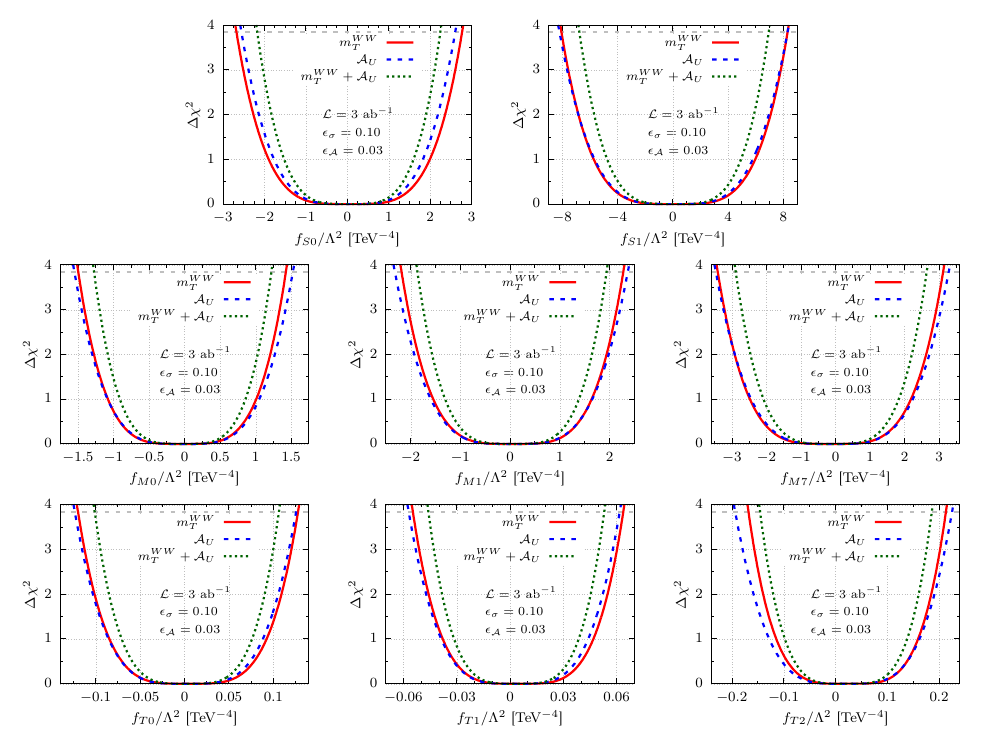}
\caption{\label{fig:comparecut}
One-parameter $\Delta \chi^2$ using the  asymmetries ($\mathcal{A}_U$)  and binned $m_T^{WW}$ distribution as a function of the  WCs  for  an integrated luminosity of $3$~ab$^{-1}$.  Systematic uncertainties of $0.1$ and $0.03$ are applied to the $m_T^{WW}$ bins and asymmetries, respectively.}
\end{figure*}
%%%%%%%%%%%%%%%%%%%%%%%%%%%%%%%%%%%%%%%%%%%%%%%%%%%%%%%%%%%%%%%%%%%%%

We then combine the $\Delta\chi^2$ functions of the ten most sensitive asymmetries, denoted collectively as ${\cal A}_U$, and compare the resulting constraints with those obtained using all 44 asymmetries. We can see from  Fig.~\ref{fig:sensitivity} 
that the 95\% C.L. limits derived from the best ten asymmetries are approximately the same as those obtained using the full set of 44 asymmetries. This reduced set, therefore, provides a minimal yet comprehensive collection of observables sufficient to capture the leading sensitivity to all couplings of interest. Employing a smaller set of observables also helps mitigate systematic uncertainties and is more favorable from an experimental perspective. Consequently, in the remainder of our analysis, we restrict ourselves to these ten most sensitive asymmetries (hereafter simply referred to as ``asymmetries'', ${\cal A}_{U}$) for each WC.  \smallskip

%%%%%%%%%%%%%%%%%%%%%%%%%%%%%%%%%%%%%%

In addition to asymmetries, we also included the  $m_T^{WW}$ distribution in our analyses  to constrain the anomalous couplings. Following the CMS study~\cite{CMS:2020gfh}, we consider the $m_T^{WW}$  binning to be
\begin{equation*}
 m_T^{WW} \in \left[0,~350,~650,~850,~1050,~\infty\right]~ \mathrm{GeV} . 
\end{equation*}
 The cross sections in each bin are parametrized as functions of the WCs according to Eq.~(\ref{eq:cpevenfit}). The $\Delta\chi^2$ function for the $m_T^{WW}$ spectrum is constructed as 
\begin{equation}
\Delta\chi^2(\mathbf{f}) = \sum_{i=1}^5 \frac{\left(\sigma_i (\mathbf{f})-\sigma_i({\rm SM})\right)^2}{\delta\sigma_i} \;,
\end{equation}
with $\delta\sigma_i = \sqrt{\sigma_i({\rm BKG})/{\cal L}+(\epsilon_\sigma\times\sigma_i({\rm BKG}))^2}$. Here, $\epsilon_\sigma$ is the systematic uncertainty in the cross section.  \smallskip

In order to validate our analysis framework, we extracted the 95\% C.L. limits on the WCs using the $m_T^{WW}$ distributions for an integrated luminosity of 137~fb$^{-1}$, assuming a systematic uncertainty of  11\%, as used in the CMS study~\cite{CMS:2020gfh}. We find that our limits are comparable to, and in some cases slightly stronger than, those reported in the CMS study~\cite{CMS:2020gfh}, primarily due to the higher center-of-mass energy considered in our analysis. \smallskip

In Fig.~\ref{fig:comparecut}, we show the one-parameter $\Delta\chi^2$ distributions as functions of the WCs 
%for an integrated luminosity of $3$~ab$^{-1}$, 
obtained using the angular asymmetries ($\mathcal{A}_{U}$), the $m_T^{WW}$ distribution and their combination. Systematic uncertainties of 0.03 and 0.10 are assigned to the asymmetries and the $m_T^{WW}$ bins, respectively. Our results show that  asymmetries alone lead to limits comparable to those obtained from the $m_T^{WW}$ analysis. Combining the asymmetries with the $m_T^{WW}$ distribution leads to an improvement in the constraints on the WCs compared to using either observable individually. %\smallskip
The corresponding 95\% C.L. limits on the WCs are presented in Table~\ref{tab:limitoneparadifflumi}
for integrated luminosities of 300, 1000, and 3000~fb$^{-1}$.
The combined limits shown in the last column improve upon those obtained from the $m_T^{WW}$ distributions alone in the second-last column by 13--20\%. \smallskip

%%%%%%%%%%%%
\begin{table*}[!htb]
\centering
\caption{\label{tab:limitoneparadifflumi}
One-parameter limits (TeV$^{-4}$) at $95\%$ C.L. on the WCs using the $m_T^{WW}$ distributions and the combination of $m_T^{WW}$ distributions and the best ten sensitive  asymmetries $(\mathcal{A}_{U})$  for integrated luminosities of 300, 1000, and 3000 fb$^{-1}$ for  systematic uncertainties on $m_T^{WW}$ distributions ($\epsilon_\sigma$) and the asymmetries ($\epsilon_{\cal A}$) to be 0.1 and 0.03, respectively  at the 13.6 TeV LHC.
}
 \renewcommand{\arraystretch}{1.5}
 \begin{tabular*}{1\textwidth}{@{\extracolsep{\fill}}lcccccc@{}}\hline
  
  Luminosity & \multicolumn{2}{c}{ 300 fb$^{-1}$}& \multicolumn{2}{c}{1000 fb$^{-1}$} & \multicolumn{2}{c}{3000 fb$^{-1}$}  \\ \hline
  Variables & $m_T^{WW}$ & $m_T^{WW}+\mathcal{A}_U$ & $m_T^{WW}$ & $m_T^{WW}+\mathcal{A}_U$ & $m_T^{WW}$ & $m_T^{WW}+\mathcal{A}_U$ \\ \hline
$f_{S0}/\Lambda^4$&$[-4.32 ,~+4.42 ]$&$[-3.69 ,~+3.76 ]$&$[-3.29 ,~+3.39 ]$&$[-2.73 ,~+2.8  ]$&$[-2.68 ,~+2.77 ]$&$[-2.17 ,~+2.24 ]$\\
$f_{S1}/\Lambda^4$&$[-13.0 ,~+13.3 ]$&$[-11.4 ,~+11.6 ]$&$[-9.91 ,~+10.2 ]$&$[-8.48 ,~+8.67 ]$&$[-8.06 ,~+8.31 ]$&$[-6.76 ,~+6.95 ]$\\
$f_{M0}/\Lambda^4$&$[-2.38 ,~+2.31 ]$&$[-2.11 ,~+2.06 ]$&$[-1.83 ,~+1.76 ]$&$[-1.59 ,~+1.53 ]$&$[-1.5  ,~+1.43 ]$&$[-1.27 ,~+1.22 ]$\\
$f_{M1}/\Lambda^4$&$[-3.53 ,~+3.63 ]$&$[-3.14 ,~+3.22 ]$&$[-2.69 ,~+2.78 ]$&$[-2.34 ,~+2.42 ]$&$[-2.18 ,~+2.28 ]$&$[-1.86 ,~+1.94 ]$\\
$f_{M7}/\Lambda^4$&$[-5.39 ,~+5.09 ]$&$[-4.75 ,~+4.5  ]$&$[-4.15 ,~+3.85 ]$&$[-3.58 ,~+3.33 ]$&$[-3.41 ,~+3.12 ]$&$[-2.89 ,~+2.64 ]$\\
$f_{T0}/\Lambda^4$&$[-0.195,~+0.203]$&$[-0.171,~+0.176]$&$[-0.148,~+0.156]$&$[-0.127,~+0.132]$&$[-0.12 ,~+0.128]$&$[-0.101,~+0.106]$\\
$f_{T1}/\Lambda^4$&$[-0.091,~+0.100]$&$[-0.079,~+0.087]$&$[-0.068,~+0.077]$&$[-0.058,~+0.065]$&$[-0.055,~+0.064]$&$[-0.046,~+0.053]$\\
$f_{T2}/\Lambda^4$&$[-0.282,~+0.328]$&$[-0.255,~+0.296]$&$[-0.211,~+0.256]$&$[-0.186,~+0.227]$&$[-0.168,~+0.213]$&$[-0.146,~+0.186]$\\
\hline
\end{tabular*}
\end{table*}
%\end{comment}
%

%%%%%%%%%%%%%%%%%%%%%%%%%%%%%%%%%%%%%%%%%%%%%%%%%%%%%%%%%%%%%%
 \begin{figure*}[!htb]
    \centering
    \includegraphics[width=1\textwidth]{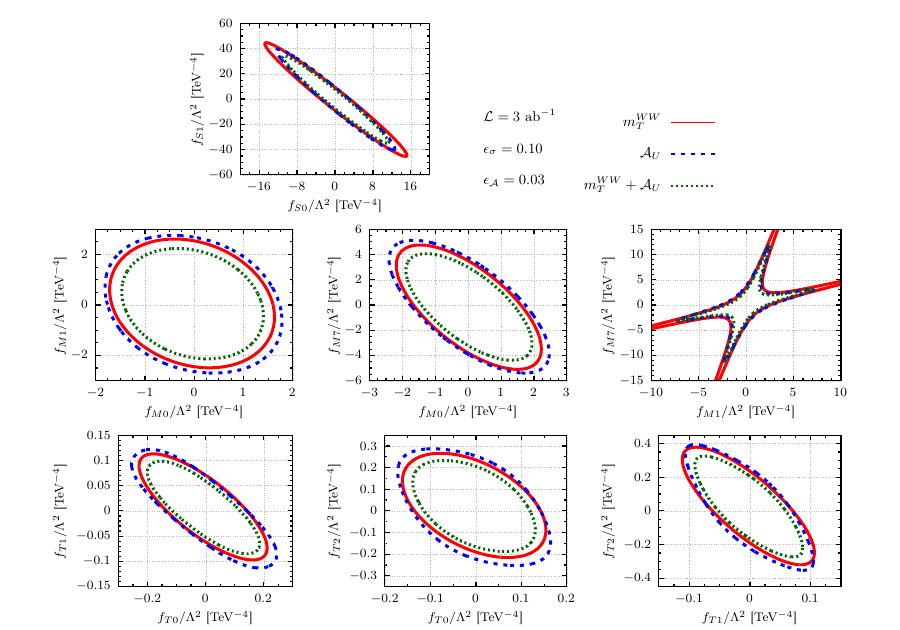}
    \caption{
   Two-parameter $95\%$ C.L. contours obtained using   asymmetries,  $m_T^{WW}$ distribution and their combination for an integrated luminosity of $3$~ab$^{-1}$. Systematic uncertainties of $0.1$ and $0.03$ are applied to the $m_T^{WW}$ bins and asymmetries, respectively.}
    \label{fig:twodw}
\end{figure*}
%%%%%%%%%%%%%%%%%%%%%%%%%%%%%%%%%%%%%%%%%%%%%%%%%%%%%%%%%%%%%%

%aqui

Next, we explore scenarios where  two anomalous WCs are non-vanishing  while  all others are set to zero. Fig.~\ref{fig:twodw} depicts the two-dimensional allowed regions at 95\% C.L. obtained from asymmetries, transverse mass distribution and their combination for an integrated luminosity of 3~ab$^{-1}$. The top panel of this figure  shows a strong anti-correlation between $f_{S0}$ and $f_{S1}$ that originates from the large interference between the $f_{S0}$ and $f_{S1}$ amplitudes as can be seen in Table~\ref{tab:cs-wc}. In this case,  ${\cal A}_U$ leads to slightly tighter constraints than from $m_T^{WW}$. From the middle-right panel we can see that  the $f_{M1}$--$f_{M7}$ plane exhibits four preferred directions in which both couplings can attain large values simultaneously, {\em i.e.} there are approximated blind directions. This is due to a sizable negative interference between the contributions of these couplings; see Table~\ref{tab:cs-wc}. Notice that the asymmetries leads to more restrictive bounds along the ``blind'' directions.  For other   WC pairs, such as $f_{M0}$--$f_{M1}$, $f_{M0}$--$f_{M7}$, and the various pairs $f_{T_i}$--$f_{T_j}$, the correlations between the couplings are relatively mild.  In all cases, the combination of the angular asymmetries and the $m_T^{WW}$ distribution yields the most stringent constraints, producing tighter exclusion contours than those obtained from either observable alone. \smallskip

%%%%%%%%%%%%%%%
\begin{figure*}[!t]
\centering
\includegraphics[width=1\textwidth]{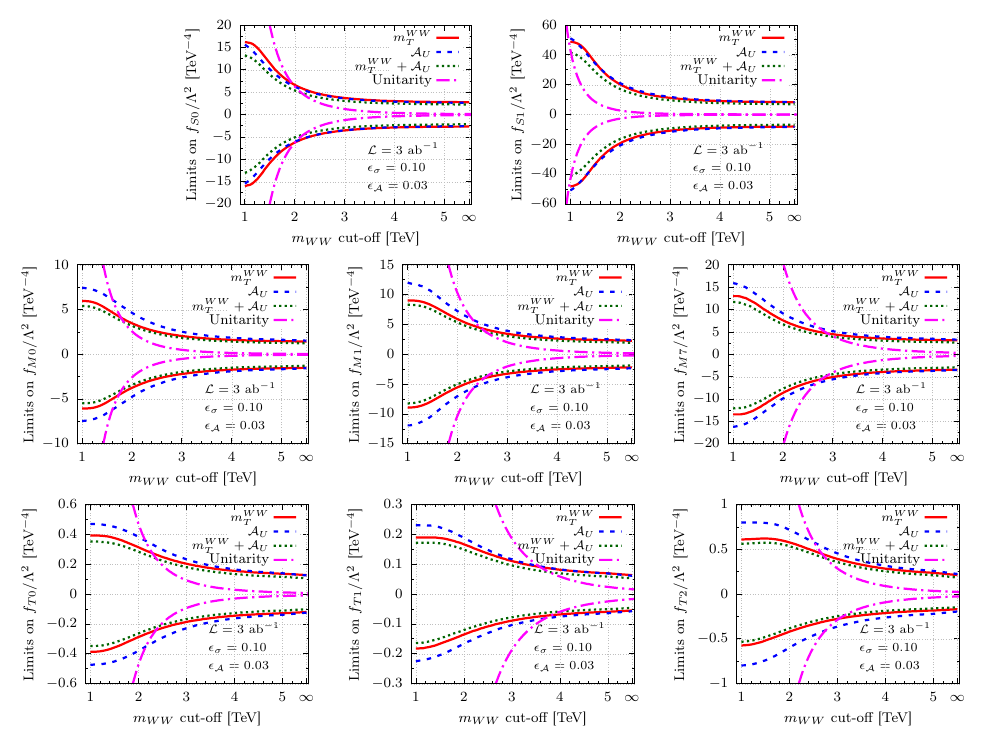}
\caption{\label{fig:lim_with_uni}
 One-parameter 95\% C.L. intervals for  just one non-vanishing WC as a function of the $WW$ invariant mass cut-off. Convention are the same as in Fig.~\ref{fig:twodw}.  The unitarity limits from Ref.~\cite{Almeida:2020ylr} are also shown.
}
\end{figure*} 
%%%%%%%%%%%%%%%%%%%%%%%%%%

In the EFT description, the presence of aQGCs causes the scattering amplitudes to grow with energy, eventually leading to a breakdown of tree-level unitarity at sufficiently high scales~\cite{Perez:2018kav,Almeida:2020ylr}. To obtain reliable and physically consistent bounds, it is therefore necessary to suppress EFT contributions in the kinematic region where unitarity is violated, while keeping the SM contribution unchanged over the full range of the $WW$ invariant mass. Next, we analyze the impact of cutting off the $WW$ invariant mass on the QGC constraints.  In this analysis, we obtain  limits on individual QGC  coefficients  excluding EFT contributions above a chosen invariant-mass threshold, denoted by $m_{WW}$ cut-off. This selection is applied at the parton level, prior to parton showering, using the kinematic information of the $WW$ system, and is implemented directly on the invariant mass $m_{WW}$.\smallskip

In Fig.~\ref{fig:lim_with_uni}, we present the one-parameter 95\% C.L. intervals on the WCs 
as functions of the $m_{WW}$ cut-off  for an integrated luminosity of $3$~ab$^{-1}$. For comparison, we also show the limits obtained without imposing any $m_{WW}$ cut-off, denoted by $\infty$ on the x-axis. Systematic uncertainties of 0.03 and 0.1 are assigned to the asymmetries and the $m_T^{WW}$ bins, respectively. The corresponding unitarity bounds from Ref.~\cite{Almeida:2020ylr} are overlaid for reference.\smallskip

We can see from Fig.~\ref{fig:lim_with_uni} that, as expected, the constraints on the WCs become weaker as the $m_{WW}$ cut-off is lowered, since an increasing fraction of the high-energy QGC contributions is removed. Furthermore,  we learn that
the EFT is reliable for $m_{WW}$ cut-offs $\lesssim 1$ TeV for  $f_{S1}$, hence weakening the bound on this WC by a factor of $\sim 5$. The maximum allowed cut-off for the other QGC WC is higher: for $f_{S0}$ and  $f_{M0}$ it is near 2~TeV, whereas for the other $f_M$ coefficients it lies in the range of 2--3~TeV, and exceeds 3~TeV for $f_{T1}$, consequently having a smaller impact on the attainable bounds on QGCs. We present in Table~\ref{tab:unitarity_safe_limits} the 95\% C.L. limits on the QGCs for the maximum $m_{WW}$ value allowed by unitarity.

%%%%%%%%%%%%%%%%%%%%%%%%%%%%%%%%%%%%%%%%%%%%%%%%%%%%%%%%%%%%%%%%%%%%%%%
\begin{table}[!htb]
\caption{  Maximum $m_{WW}$ cut-off and corresponding 95\% C.L. limits on WCs for an integrated luminosity of 3 ab$^{-1}$.  
}
\label{tab:unitarity_safe_limits}
\centering
\renewcommand{\arraystretch}{1.5}
\begin{tabular*}{0.45\textwidth}{@{\extracolsep{\fill}}lcc@{}}\hline
WC   & $m_{WW}$ cut-off (TeV) & Limit (TeV$^{-4}$) \\ \hline
%& for lower limit & for higher limit (remove) &  \\ \hline
$f_{S0}/\Lambda^4$&     $2.2    $ &$[-4.37       ,~+4.76 ]$\\
$f_{S1}/\Lambda^4$&     $1.0    $&$[-41.1       ,~+41.5 ]$\\
$f_{M0}/\Lambda^4$&     $1.8    $&$[-3.94       ,~+3.56 ]$\\
$f_{M1}/\Lambda^4$&     $2.65   $&$[-3.35       ,~+3.67 ]$\\
$f_{M7}/\Lambda^4$&     $2.9    $&$[-4.62       ,~+3.87 ]$\\
$f_{T0}/\Lambda^4$&     $2.45   $&$[-0.211      ,~+0.241]$\\
$f_{T1}/\Lambda^4$&     $4.0    $&$[-0.058      ,~+0.075]$\\
$f_{T2}/\Lambda^4$&     $3.15   $&$[-0.234      ,~+0.4  ]$\\\hline
\end{tabular*}
\end{table}
%%%%%%%%%%%%%%%%%%%%%

%%%%%%%%%%%%%%%%%%%%%%%%%%%%%%%%%%%%%%%%%%%%%%%%%%%%%%%%%%%%%%%%%%%%%%%%%%%%%
\section{Conclusion}
\label{sec:conclude}
In this work, we analyzed the potentiality of the  same-sign $W$ pair production by VBS at the HL-LHC to probe genuine anomalous  quartic gauge couplings. In addition to the canonical study of the transverse mass of  $W$ pairs we also took into account polarization and spin correlation parameters. We focused on the   fully leptonic decays of the same-sign $W$ pair which contain two neutrinos that escape detection.  We reconstructed the  neutrino momenta  using multi-layer neural network, which allowed us obtaining  the $W$-boson momenta with reasonable accuracy. We further employed the joint angular distribution of final decayed leptons in the rest frame of $W$ bosons to obtain the parameters of the $WW$ density matrix. These parameters are obtained as asymmetries in the angular functions of decayed leptons.  \smallskip

The polarization and correlation parameters depend on the quartic $W$ vertex, thus, any anomalous shift in the vertex can be probed with asymmetries of angular functions. An additional observable in the form of transverse mass of the $W^\pm W^\pm$ pair was also included in the analyses. Moreover, to reduce the uncertainty in measuring observables, we worked with just the best ten sensitive asymmetries. We learned that  the limits on WCs  using only the best ten sensitive asymmetries are  comparable to those obtained using all $44$ CP-even asymmetries. A combined analysis incorporating both angular asymmetries and kinematic information yields tighter constraints than those obtained from either approach individually. \smallskip

We also examined the impact of unitarity constraints by imposing an upper cut on the invariant mass of the reconstructed diboson system in the QGC contribution. We studied the dependence of the limits on the Wilson coefficients by varying the $m_{WW}$ cut-off over a wide range and identified the corresponding unitarity-safe regions. We found that the constraints derived from both angular asymmetries and the binned $m_T^{WW}$ distribution are affected by the choice of the $m_{WW}$ cut-off, with the limits generally becoming weaker as more high-energy EFT contributions are removed. This behavior reflects the loss of sensitivity induced by unitarity-motivated kinematic restrictions. \smallskip

Our results clearly indicate that including the polarization and correlation asymmetries in the analysis enhances the LHC potential to study anomalous quartic gauge couplings. Ergo, the use of polarization and spin-correlation observables in this process is  relevant not only for future LHC runs but also for existing datasets, particularly in view of the recent evidence for polarized $W$ bosons in same-sign vector boson scattering reported by the ATLAS Collaboration~\cite{ATLAS:2025wuw}.

%%%%%%%%%%%%%%%%%%%%%%%%%%%%%%%%%%%%%%%%%%%%%%%%%%%%%%%%%%%%%%%%%%%%%%
\section*{Acknowledgments}

AS is grateful to Ritesh K. Singh for many valuable discussions and helpful suggestions. 
AS  acknowledges support from the National Natural Science Foundation of China under Grant Nos. T2241005 and 12075059. RR is supported by FAPESP fellowship with grant 2023/04036-1 and  2025/06648-0 . OJPE is partially supported by the CNPq grant number 302120/2025-4.

\bibliography{refer2}

\end{document}